# Photon Absorption Remote Sensing (PARS): Comprehensive Absorption Imaging Enabling Label-Free Biomolecule Characterization and Mapping

Benjamin R. Ecclestone[1], James A. Tummon Simmons[1], James E.D. Tweel[1], Deepak Dinakaran[2,3], Parsin Haji Reza[1, *]

[1] *PhotoMedicine Labs, University of Waterloo, 200 University Ave W, Waterloo, ON N2L 3G1, Canada.*
[2] *Sunnybrook Research Institute, University of Toronto, 2075 Bayview Avenue, Toronto, ON M4N, 3M5, Canada.*
[3]*Medical Biophysics, Temerty Faculty of Medicine, Toronto, ON, 2075 Bayview Avenue, Toronto, ON, M4N 3M5, Canada.*
*Corresponding author: phajireza@uwaterloo.ca

**Abstract** – *Label-free optical absorption microscopy techniques continue to evolve as promising tools for label-free histopathological imaging of cells and tissues. However, critical challenges relating to specificity and contrast, as compared to current gold-standard methods continue to hamper adoption. This work introduces Photon Absorption Remote Sensing (PARS), a new absorption microscope modality, which simultaneously captures the dominant de-excitation processes following an absorption event. In PARS, radiative (auto-fluorescence) and non-radiative (photothermal and photoacoustic) relaxation processes are collected simultaneously, providing enhanced specificity to a range of biomolecules. As an example, a multiwavelength PARS system featuring UV (266 nm) and visible (532 nm) excitation is applied to imaging human skin, and murine brain tissue samples. It is shown that PARS can directly characterize, differentiate, and unmix, clinically relevant biomolecules inside complex tissues samples using established statistical processing methods. Gaussian mixture models (GMM) are used to characterize clinically relevant biomolecules (e.g., white, and gray matter) based on their PARS signals, while non-negative least squares (NNLS) is applied to map the biomolecule abundance in murine brain tissues, without stained ground truth images or deep-learning methods. PARS unmixing and abundance estimates are directly validated and compared against chemically stained ground truth images, and deep learning based-image transforms. Overall, it is found that the PARS unique and rich contrast may provide comprehensive, and otherwise inaccessible, label-free characterization of molecular pathology, representing a new source of data to develop AI and machine learning methods for diagnostics and visualization.*

## I. Introduction

Over the past century, visualizing the microanatomy of cells and tissues through chemical staining of subcellular tissue structures has shaped histopathology, clinical oncology, and medical diagnostics. Innovations in chemical labelling, such as immunohistochemical and molecular specific staining, have transformed biological understanding, and oncological processes by revolutionizing malignancy subtyping and tumor classification.[1,2] Unfortunately, contrast agents unavoidably interfere with specimens' physical and chemical integrity, [3–7] meaning samples are effectively consumed once stained. This prohibits multiple stains from being applied to a single specimen. [3–7] As a direct outcome, the volume of diagnostic tissue specimens can severely limit the number of contrasts and diagnostic tests, especially in clinical settings where there are limited sample volumes (e.g., brain tumor biopsies).[3–7]

In contrast, label-free imaging permits visualization of the microanatomy and biochemistry of living and preserved cells and tissues without exogenous labels and chemical modification.[8,9] Optical absorption and non-linear scattering interactions especially, are largely driven by chemical structure and composition, directly enabling non-invasive label-free structural [9–11], functional,[12,13] and molecular imaging.[14,15] To this end, there are two major categories of modalities which exhibit potential for label-free tissue imaging: (1) linear electronic absorption methods including autofluorescence,[16–19] photothermal,[20–22] and photoacoustic[23–26] microscopy (2) non-linear techniques like stimulated Raman scattering (SRS),[27–291] multiphoton,[30,31] and harmonic generation.[30,32–34] In recent publications, these label-free absorption and non-linear scattering methods have been explored for applications including label-free direct characterization of tissue malignancy, [28,35] intraoperative imaging of brain tumor tissues, [28,36] and label-free virtual histochemical staining. [15,16,37,38]

To this end, linear absorption methods (e.g., autofluorescence) have begun to gain traction towards early adoption for label-free tissue imaging and histopathology applications (e.g., virtual histochemical staining), as the contrast is easy to capture across a wide range of specimens (e.g., linear cross sections ($\sim 10^{-15}$ to $10^{-20} cm^2$)[39] are orders of magnitude higher than non-linear cross sections ($\sim 10^{-25}$ to $10^{-33} cm^2$)). [40,41] Linear absorption modalities normally provide contrast by inducing excited state transition, then observing either relaxation or de-excitation effects. De-excitation from the excited state occurs through two main processes, radiative and non-radiative transitions, where the division of radiative vs. non-radiative relaxation is quantified by the fluorescence quantum yield (QY). [42] During non-radiative relaxation, energy is shed to the surrounding media through collisions or vibrations, over the scale of picoseconds [43,44]. During radiative relaxation, absorbed energy is released through the emission of photons over a period of pico- to nano- seconds, depending on the molecules excited state lifetime [44]. Prevalent absorption modalities usually target contrast from one of these de-excitation effects and are often grouped accordingly as either radiative (autofluorescence, fluorescence lifetime microscopy) [44,45] or non-radiative (photoacoustic,[46,47] and photothermal[21,48]) methods.

Here we introduce Photon Absorption Remote Sensing (PARS), an emerging pump-probe method which is designed to better leverage the multitude of contrasts afforded by an absorption event.[49–53] Unlike prevalent absorption modalities which target either radiative (autofluorescence) or non-radiative (photothermal and photoacoustic) effects, PARS aims to capture a complete representation of the relaxation phenomena. Following excitation, PARS uses a confocal probe laser to detect non-radiative (photothermal and photoacoustic) induced material changes and optical modulations, while radiative emissions are measured by capturing emitted photons. Simultaneously measuring aspects of both dominant de-excitation pathways (radiative and non-radiative) can circumvent mechanism specific contrast limitations, such as the fluorescence quantum yield. Any biomolecule which absorbs light will offer some degree of PARS contrast (either radiative or non-radiative) enabling unique insights. For instance, PARS can readily image low quantum yield (QY) biomolecules, (e.g., nuclei and hemeproteins) [54,55] and high QY connective tissues, [25,56] which may be difficult or impossible to capture with independent radiative (e.g. autofluorescence) or non-radiative (e.g. photoacoustic) methods, respectively. The combination of these absorption fractions, presented as the "Total-Absorption" (TA) mapping, indicates the absolute level of absorption. The ratio of relaxation fractions, proposed as the quantum efficiency ratio (QER), indicates the quantum efficiency characteristics of a biomolecule.[53]

This work (1) Presents the first comprehensive explanation of the PARS mechanism, (2) Illustrates the optical architecture and corresponding methods used to characterize the radiative (emission intensity), and non-radiative (energy and temporal evolution) at each excitation event. (2) Demonstrates direct characterization, differentiation, and unmixing of biomolecules label-free inside complex samples, through statistical processing of the rich PARS measurements.

A multiwavelength PARS system is presented featuring 266 nm and 532 nm excitation for label-free imaging of tissues. The presented microscope characterizes radiative (emission intensity), and non-radiative (energy and temporal evolution) relaxation at both (266 nm and 532 nm) excitation wavelengths. As an example, the PARS system is applied to imaging thin unstained sections of formalin fixed paraffin embedded (FFPE) human skin and murine brain tissues mounted on slides, which are an intermediate format generated during chemical staining-based histopathology preparation. Through this process PARS seamlessly integrates into histochemical workflows, capturing label-free high-resolution whole slide images of entire tissues specimens at the equivalent of 40x digital scanning, while allowing downstream chemical staining of the same tissue sections. Previously, this one-to-one PARS imaging and chemical staining workflow was applied to develop deep learning based virtual hematoxylin and eosin (H&E) images which were shown to be indistinguishable from, and diagnostically equivalent to chemical H&E staining. [51,57]

This work directly explores the unique array of measurements captured in PARS to demonstrate capacity to characterize, differentiate, and unmix clinically relevant biomolecules label-free inside complex samples, without applying black-box virtual staining or deep-learning methods. Established statistical processing methods (Gaussian mixture models), are used to perform masked endmember extraction developing characteristic PARS signatures for endogenous biomolecules. Resulting PARS signatures are used to unmix biomolecule abundances in complex specimens through non-negative least squares (NNLS) unmixing. Applied to unstained brain tissue samples, PARS provides direct specificity and abundance mapping for critical biomolecules including nuclei, red blood cells, myelinated (white matter) and unmyelinated neurons (gray matter). Biomolecular mappings are compared and validated against chemical H&E, and DAPI ground truth images. Finally, the presented statistical method is applied to develop H&E-like visualizations in brain tissues, directly circumventing black-box deep-learning methods.

Overall, PARS emerges as a highly promising high-resolution comprehensive absorption modality, offering powerful label-free visualizations within the presented FFPE tissue specimens, that may be otherwise inaccessible with conventional imaging techniques. Concurrently, the presented GMM & NNLS workflow is shown to reliably characterize and map diagnostic biomolecules facilitating label-free histology, directly avoiding the requirement for ground truth chemical stains. The presented results illustrate PARS unique potential to drive the development of robust deep-learning methods for virtual staining, and diagnostics in future.

## II. PARS Mechanism

To understand the guiding principle behind PARS, consider the interactions which occur as a pulse of light is absorbed by biomolecules (Figure 1). During electronic absorption (Excitation), molecules will capture the energy of incident photons resulting in excited state transition.[21,44,48] De-excitation subsequently occurs through radiative (fluorescence) relaxation and non-radiative (thermal) relaxation, where the division of de-excitation is characterized by the fluorescence yield (QY).[42] Fluorescence or radiative relaxation results in the release of energy through Stokes shifted auto-fluorescent photons, while non-radiative relaxation results in the release of energy through heating of the surrounding media via collisions or vibrations [21,44,48]. PARS induces excitation, then observes the temporal evolution, intensity, and distribution of these de-excitation phenomena to characterize absorbers using a pump-probe architecture (Figure 1).

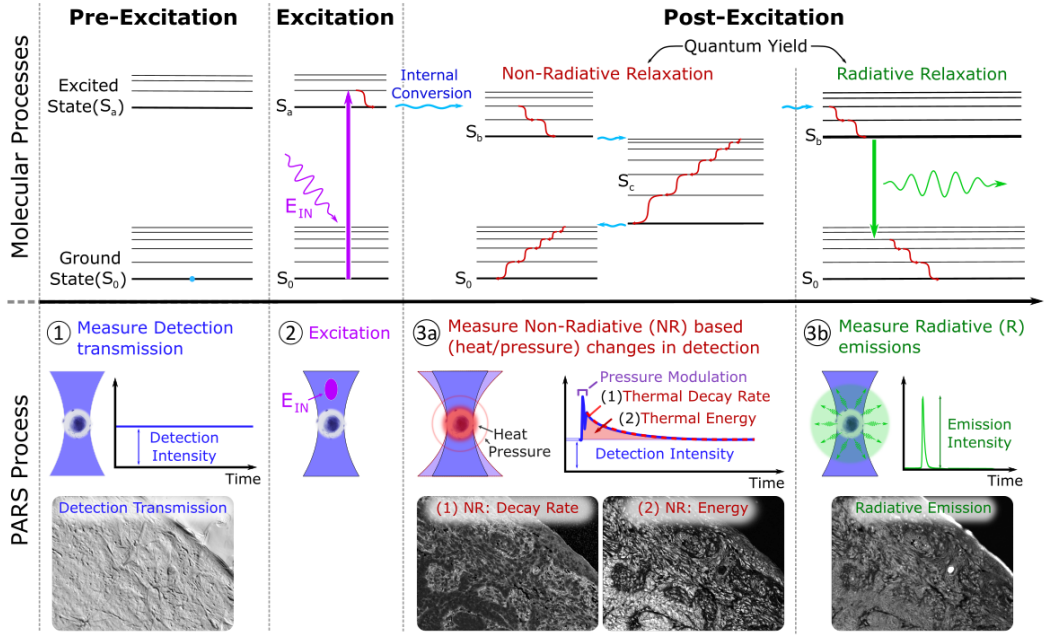

**Figure 1:** Electronic excitation and relaxation processes experienced within a biomolecule, compared to the measurements used in PARS to characterize the photophysical processes delivering absorption contrast. Absorption events are characterized as follows. (1) First a baseline measurement of the detection transmission/reflection is measured. (2) Next excitation is induced, causing transition from the ground state ($S_0$) to an excited state (e.g., $S_a, S_b, S_c$). (3) The de-excitation processes are measured to characterize the relaxation intensities, temporal evolution, and distribution (i.e., quantum yield). (3a) Non-radiative relaxation is characterized by measuring transient heating induced modulations in the local optical properties, expressed as transmissivity modulations in the detection beam. In this example, the non-radiative signals are characterized based on modulation energy, and temporal evolution. (3b) Radiative emissions are directly quantified as the emission of Stokes shifted photons.

In practice the PARS imaging process consists of three sequential steps (Figure 1). (1) First, baseline optical properties are measured by characterizing the transmission (or reflection) of the detection laser. (2) Next the co-focused pump (excitation) laser is pulsed inducing excited state transition. (3) Finally, the transient relaxation effects are measured to characterize the absorption event. (3b) Quantifying radiative relaxation intensity is relatively straightforward since radiative transitions are measured as the emission of Stokes shifted auto-fluorescent photons (analogous to autofluorescence microscopy). In this embodiment, radiative photons are spectrally isolated and directed to a photodiode. Peak emission intensity is measured as the Radiative relaxation amplitude ($R_a$) and directly used as a pixel intensity value. (3a) In contrast measuring non-radiative relaxation presents inherent complexities, as the transitions cannot be directly measured. Instead, non-radiative de-excitation is quantified by using the confocal detection beam to observe secondary effects caused by the localized non-radiative induced heating.

Under normal circumstances, the main consequence of non-radiative (photothermal) heating is localized thermo-elastic expansion.[21,48,58] In some cases, if heating is sufficiently rapid (e.g., on the scale of nano- to pico-seconds), thermal expansion will outpace relaxation creating appreciable (photoacoustic) pressures.[47,59] The photothermal and photoacoustic modulations will quickly propagate within samples, generating transient strain and perturbation in sample density. In turn, these effects will modulate the local optical properties in several ways, such as refractive index changes (due to density variations),[58,60–63] surface deformation,[64] and sample deflection (due to strain).[65,66]

The temporal evolution of these pressure and temperature modulations are subsequently coupled to specimens' material properties.[67] Photoacoustic signals will travel at the speed of sound according to wave propagation, meaning the perturbation lifetime will be dictated by the period of the generated pressure waves. Using photoacoustic principles, the period of an excited wave is approximated through the photoacoustic confinement time, calculated as $[\tau_p = 2w/v_s]$.[68] In this equation, $w$ is the absorber diameter, and $v_s$ is the speed of sound in the surrounding media. Thermal signals will evolve and propagate according to thermal diffusion, as the deposited heat will dissipate following Newtons law of cooling. This is characterized by an exponential decay, where the local temperature $T$ will dissipate proportional to: $[T \propto \exp(-\tau_t t)]$, where $[\tau_t = w^2/4k]$.[21] In this equation $k$ is the thermal diffusivity, and $w$ is the absorber diameter. Approximately a 98% decay in heat (and signal intensity) will occur across a time of $4\tau_t$.

As an example, the impulse response of the photoacoustic and photothermal signals are calculated for the FFPE tissues imaged in this work. For this example, it is assumed that the FFPE tissues are a semi-infinite absorber, with material properties equivalent to paraffin wax, as the water in the tissues has been replaced with paraffin. Assuming an absorber with a diameter identical to the excitation spot size ($w = 850\ nm$), the photoacoustic lifetime ($\tau_p = 2w/v_s$) is calculated to be $\tau_p \approx 1.6\ ns$ in paraffin wax with speed of sound $v_s = 1050\ m/s$.[69] The thermal lifetime ($\tau_t = w^2/4k$) is calculated to be $\tau_t \approx 1.2\mu s$, with a 98% decay over $4\tau_t \approx 4.8\mu s$, assuming $k = 0.15 mm^2 \cdot s^{-1}$, and $w = 850\ nm$.[70]

With the excitation and non-radiative detection laser co-focused, these photothermal and photoacoustic perturbations are observed from the moment of generation until they propagate out of the detection focal spot. The perturbation intensity is directly quantified by measuring changes in the reflected or transmitted detection intensity relative to an unexcited baseline. Resulting non-radiative signals will consist of a superimposed mixture of both photothermal and photoacoustic effects.[71] However, as the impulse response calculations show, the respective modulations evolve at different orders of speed. Signals can subsequently be broken into two corresponding regimes, where modulations will be a combination of pressure and temperature effects across short time scales (pico- to nano-seconds), and mostly thermal effects over the longer term (nano-seconds and on).

Depending on the embodiment PARS may target and viewed different aspects of the pressure and temperature modulations.[67] For example, recent works have explored characterizing absorber size by assessing the photoacoustic lifetime $(\tau_p)$ of the non-radiative signals.[67] In the PARS embodiment and thin sectioned histology specimens explored in this work, the non-radiative signals are expected to be largely dominated by the thermal contribution as photoacoustic transit time is very short ($\sim 1.6\ ns$) compared to the ($\sim 500\ ns$) time domain signal recording. Hence, in this application, two values are extracted from each non-radiative signal (Figure 1) based on the (thermal) Newtons cooling model, (1) signal energy ($NR_E$) calculated as the modulation integral, and (2) the signals thermal decay rate ($NR_D$) calculated as the log-linear slope of the exponential decay signal. The signal energy aims to quantify the total level of non-radiative de-excitation, where the modulation energy will be correlated to the temperature change and level of non-radiative relaxation. The signal decay rate aims to characterize the thermal relaxation parameter ($\tau_T$), which is correlated to the local material properties through the thermal diffusivity.

In total, this PARS embodiment quantifies the radiative amplitude ($R_A$), non-radiative energy ($NR_E$), and non-radiative decay rate ($NR_D$) from every excitation from each wavelength of excitation used. Critically, all contrasts are measured simultaneously from the same excitation, resulting in a comprehensive representation of the photophysical phenomena (except the radiative emission spectra and lifetime) occurring during an absorption event. The array of contrasts directly leverages the strengths and advantages of both radiative (auto-fluorescence), and non-radiative (photothermal, and photoacoustic) techniques, where capturing the interplay of relaxation phenomena may also provide PARS certain advantages. For instance, in PARS, absorption is measured as the total energy shed through both de-excitation phenomena, decoupling absorption measurements from mechanism specific factors (e.g., QY). This directly mitigates a critical limitation of alternative methods, which may struggle to properly characterize biomolecules (e.g., nuclei and hemeproteins) due to their QY. Conversely, the relative distribution of de-excitation phenomena can reveal valuable insight into biomolecular characteristics such as the QY. In this work, these unique representations facilitated by PARS imaging are explored for histological imaging.

## III.    Methods

### A.    PARS System Architecture

A simplified diagram illustrating the multiwavelength PARS microscope introduced in this work is presented in Figure 2. A pair of 266 nm 400 ps pulsed 50 kHz diode lasers provide the two excitation wavelengths (266 nm and 532 nm) (Wedge XF 266, Bright Solutions). Though 266 nm is the intended laser output, 532 nm pulses may be isolated from leftovers/leakage attributed to the harmonic generation process. The non-radiative detection is a continuous wave Coherent Obis 405 nm (OBIS-LS 405, Coherent). The output energy of each laser source is monitored by using a beam sampler (BS1: BS028 Thorlabs, BS2: BSF10-A Thorlabs, BS3: BSF10-UV Thorlabs) to redirect a small portion of the laser output to a photodiode (PD1: APD130A2 Thorlabs) for measurement.

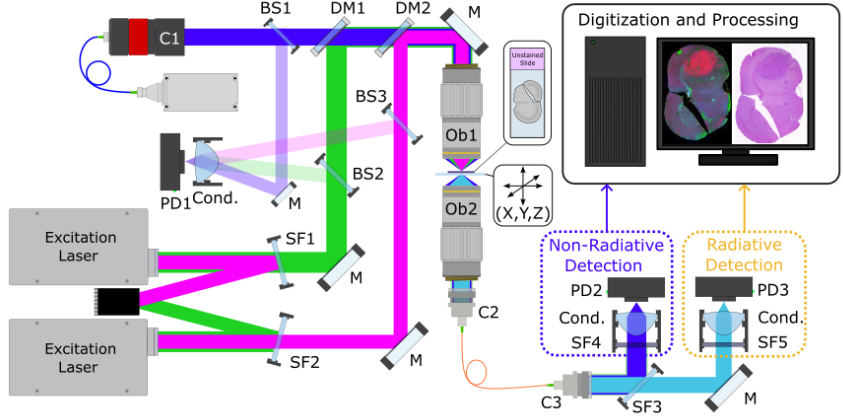
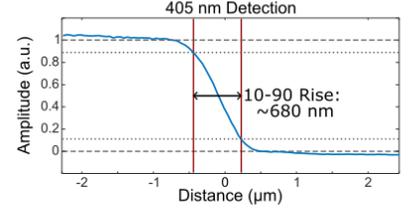
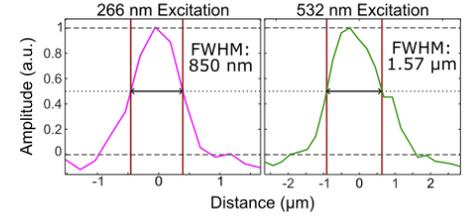
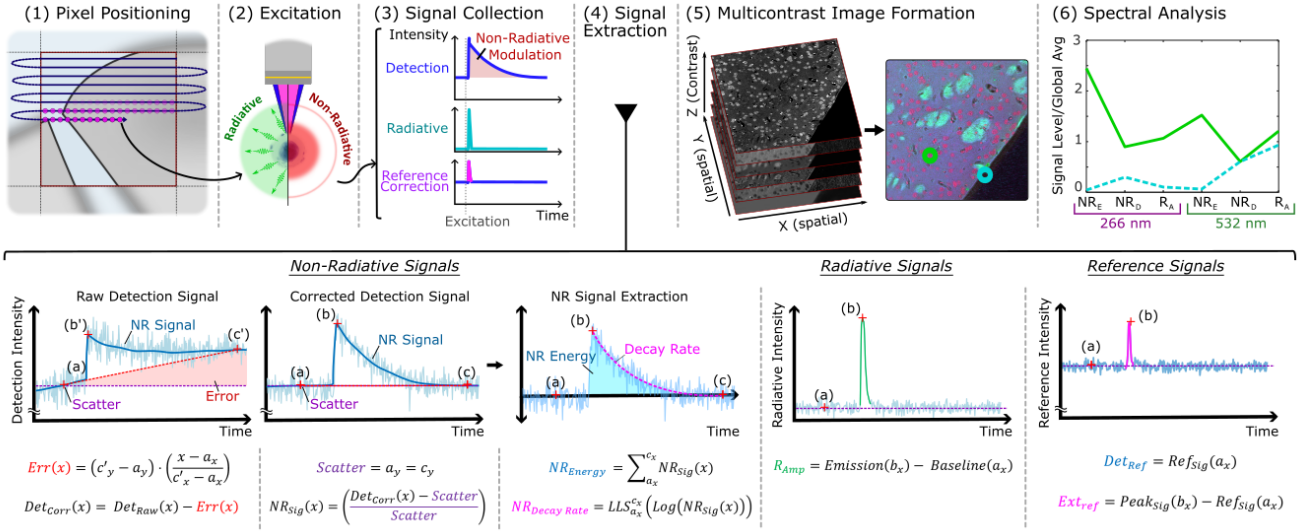

**Figure 2:** Simplified system diagram, imaging characterization, and image formation methods. (a) Simplified system diagram of the PARS microscope presented in this work. System components are as follows: SF (Spectral Filter), M (Mirror), Cond. (Condenser Lens), PD (Photodiode), Obj. (Objective Lens), DM (Dichroic Mirror), and BS (Beam Sampler). (b) Detection laser and non-radiative resolution measured from an edge spread function generated on a USAF resolution target. (c) Radiative imaging resolutions at 266 nm and 532 nm excitation wavelength, measured from the PSF generated on sub-resolution fluorescent emitters. (d) Imaging workflow showing the steps for collecting PARS data, including pixel positioning, signal processing/extraction, and resulting image formation. (4) Signal Processing Steps: **Non-Radiative Signals:** Detection intensity is corrected to remove Baseline error ($Err$) caused by changes in intensity before (a), and after (c') excitation. Baseline corrected signals ($Det_{corr}$) are reference corrected and normalized by the local scattering/transmissivity ($Scatter$ or $a$). Isolate and scaled non-radiative modulation ($NR_{Sig}$) are processed to extract (1) energy ($NR_{Energy}$), and (2) decay rate ($NR_{Decay}$). $NR_{Energy}$ is measured as the modulation integral, $NR_{Decay}$ calculated as the log-linear fitted modulation slope (using only post-excitation measurements (b) to (c)). **Radiative Signals:** Radiative emission amplitude ($R_{amp}$) is measured as the difference between the reference corrected peak signal amplitude and pre-excitation baseline. **Reference Correction:** Detection laser reference ($Det_{ref}$) is measured as the average intensity. Excitation laser reference ($Ext_{ref}$) is then measured as the difference between the peak intensity and the detection laser intensity.

For imaging, the detection and excitation beams are combined using a series of dichroic mirrors (DM1: DMSP505 Thorlabs, DM2: 37-721 Edmund Optics). The combined beams are co-focused using a 0.42 NA UV-VIS objective lens (Ob1:

NPAL-50-UV-YSTF OptoSigma). Absorption signals are measured on the opposite side of the sample in transmission mode using a high-NA Visible objective lens (Ob2: 278-806-3, Mitutoyo) which collects fluorescence emissions, and transmitted detection light. With the configuration used in this publication, the non-radiative resolutions were measured to be ~680 nm for both excitation wavelengths, based on the 10:90 rise, of the edge spread function captured on a USAF resolution target (R1DS1P1 Thorlabs). Radiative resolutions were measured to be ~850 nm and 1.57 µm for the 266 nm and 532 nm excitation wavelengths respectively, based on the point spread function generated from sub resolution fluorescent particles.

To extract absorption measurements from the optical signal, the transmitted detection light and radiative emissions are coupled into a multimode fiber (M133L01, Thorlabs) with a high-NA fiber coupler (F950FC-A, Thorlabs). Light from the multimode fiber is collimated and directed to a series of detection modules, which measure the absorption signals. The non-radiative detection is isolated by chromatic notch filtering (SF3: NF405-13 Thorlabs) and the intensity is measured using a detection module containing a spectral filter (SF4: FBH405-10 Thorlabs), focusing lens (Cond.: ACL25416U Thorlabs), and photodiode (PD2: APD130A2 Thorlabs). Remaining light containing the isolated Stokes shifted fluorescence emissions is measured using a similar detection module (NF5: NF533-17 Thorlabs, Cond.: ACL25416U Thorlabs, PD3: APD130A2 Thorlabs).

## B. *Image Formation*

The proposed PARS system operates in a point scanning architecture, where each excitation event corresponds to one pixel of data. To form images the sample is mechanically scanned across the objective lens in a raster pattern while the excitation lasers are pulsed continuously at 50 kHz. Scanning velocity is adjusted to provide a 250 nm spacing between each excitation event, corresponding to 250 nm pixels or 40x digital pathology equivalent. At each excitation (one image pixel), a 200 MHz digitizer (CSE1442, RZE-004-200, Gage Applied) records a position signal from the scanning stages, and ~500 ns of time resolved data from each system photodiode. The pixel location is calculated from the position signal, while the digitized photodiode outputs are processed to produce pixel values for each contrast. The specific signal extraction methods are outlined in the following *Signal Extraction* section. To form whole slide images, a series of overlapping subframes is captured covering the entire region of interest. In this embodiment, each subframe covers a 500 *µm* by 500 *µm* area corresponding to 2000 by 2000-pixels. The workflow for autofocusing, scanning, and stitching these subframes into a whole slide image is outlined in more depth by Tweel *et al.*[50]

Stitched whole slide images may be presented as either individual contrast channels (grayscale) or as a merged TA image. In the merged images, the various channels are mixed to form an RGB representation facilitating comparison between contrasts. In practice, each channel is normalized by its global average and mixed in RGB space using QuPath linear color mixing.[72] In general terms, the blending operation uses two steps. First, each contrast channel is mapped to a unique color vector, where black equates to zero signal while full saturation equates to maximum signal. Then, all channels are then summed to form an image. Resulting RGB representations may be contrast-clipped to produce a final image.

## C. *Signal Extraction*

At each excitation event, a total of four digitized signals are collected (Figure 2d: Signal Extraction): the NR photodiode (PD2), the radiative photodiode (PD3), the reference correction photodiode (PD1), and a stage position signal. The average stage position signal is computed and used to map the excitation event to a single (X,Y) pixel location on the subframe. Time-resolved photodiode data is then extracted to produce pixel values for the multi-dimensional PARS images as follows.

***Reference Correction:*** Reference correction signals (PD1) are used to normalize the absorption and scattering measurements to the input energies of the detection and excitation lasers. In this embodiment both the excitation and detection reference correction measurements are collected using the same photodiode (PD1) to conserve digitization channels. The continuous wave detection laser reference ($Det_{ref}$) is calculated as the average intensity prior to pulsing the excitation. The excitation laser reference energy ($Ext_{ref}$) is then measured as the difference between the peak intensity and the detection laser intensity (Figure 2d).

***Radiative Signals:*** Radiative emission amplitude (PD3: $R_{amp}$) is calculated by subtracting the peak signal amplitude from a pre-excitation baseline. Baseline correction helps mitigate stray light artifacts in the radiative measurements. Resulting radiative intensities are normalized by the reference excitation energy.

***Non-Radiative Signals:*** A new processing scheme is applied to extract the non-radiative signals based on a recent work by Tummon Simmons *et al.*[49] This workflow improves SNR while enhancing characterization of the non-radiative signal energy and shape. The full processing scheme is outlined in Figure 2(d): Signal Extraction: Non-Radiative Signals.

To extract characteristics of the non-radiative signals, the transient modulation must be isolated within the detection intensity measurement. The initial step is determining a baseline correction for each pixel. Several effects (e.g., transient power fluctuations, sample movement) can cause slow changes in the apparent intensity of transmitted detection light ($Det_{Raw}$). To remove these effects, the transmissivity is measured before excitation (denoted as (a)), and after the non-radiative signal has fully decayed (denoted as c')). Baseline error ($Err$) or the change in transmissivity from (a) to (c'), is then subtracted from the signal. Baseline corrected signals ($Det_{corr}$) are then normalized to the inputted optical energies and a correction is made to account for the local scattering/transmissivity ($Scatter$). The scattering correction involves subtracting and dividing the time resolved signal by the pre-excitation transmissivity (denoted as (a)). This operation serves to (1) isolate the non-radiative modulation ($NR_{Sig}$) from the scattering and (2) normalizes the signal amplitude to remove any dependency on the local transmissivity. These isolated and normalized time-resolved signals represent the non-radiative perturbations as a percentage change in the local scattering/transmissivity. Two features are extracted from these non-radiative signals (1) energy ($NR_{Energy}$), and (2) decay rate ($NR_{Decay}$). Signal energy is measured as the integrated area of the modulation. Decay rate is calculated as the slope by from a log-linear fitting of the modulation. For the log-linear fitting, only the post-excitation portion of the signal from (b) to (c) is used, where the signals are normalized by peak intensity (b). For reference, additional context on the non-radiative signals is presented in the *PARS Mechanism* section.

## D.     *Endmember Extraction: Gaussian Mixture Modelling*

***Forwards Model:*** Excited biomolecules will exhibit distinctive relaxation properties based on intrinsic factors (e.g., chemical structure, chemical bonding, QY, metabolic state). In turn, PARS will capture a unique mixture of contrasts, forming a PARS signature for each biomolecule known as an endmember. This logic can be extended to formulate a model for PARS signal composition in complex tissues, where an absorption event will excite a small volume of tissues (one voxel). Each voxel is approximated as a cylindrical volume of diameter roughly equal to the excitation focal resolution and a height equivalent to the tissue section thickness (~5 $\mu m$). The tissue may contain numerous biomolecule constituents within a voxel, where each biomolecule will contribute to the measured PARS signal. Subsequently, the PARS signal ($y_n$) for a given pixel ($n$) is posed as a linear mixture of the PARS signature or endmembers ($\mu_k$) attributed to the ($k$) underlying biomolecules. This is formulated as:

$$y_n = \sum_{k=1}^{K}(\mu_k \cdot a_{nk}) + \sigma_n$$

Where $a_{nk}$ is the abundance of each respective biomolecule constituent, while $\sigma_n$ is the measurement noise in the PARS signal at each pixel. Note, that the abundances $a_{nk}$ are subject to the positivity and sum-to-one constraints ($a_{nk} \geq 0$ and $1 = \sum_k a_{nk}$).

Given a large volume of data, the underlying linear mixture process responsible for generating the signals ($y_n$) can be modelled with a statistical distribution. Assuming the measurement noise ($\sigma_n$) for each contrast follows an independent gaussian distribution, the linear mixture model is reformulated as a stochastic gaussian mixture model (GMM).[73] For a given measurement ($y_n$) and a prior estimate on the number of potential biomolecules ($k$), the GMM describes the probability ($p_{nk}$) of each pixel being a given biomolecule which is characterized by a PARS signature or endmember $\mu_k$. Given parameters ($\mu_k, \Sigma_k, \pi_k$), the probability is described by:

$$p_{nk}(y_n|\mu_k, \Sigma_k) = \Pi_k \cdot \left((2\pi)^{\frac{D}{2}}|\Sigma_k|\right)^{-1/2} \cdot \exp\left\{-\frac{1}{2} \cdot (y_n - \mu_k)_n^T \cdot \Sigma_k^{-1} \cdot (y_n - \mu_k)\right\}$$

Where $\Pi_k$ is the probability weighting, or global prevalence of each endmember, $\Sigma_K$ is the covariance of each PARS endmember distribution, and $D$ is the dimensionality of the input PARS data and endmembers.

***Inverse Model:*** The presented mixture model can also be inverted to extract PARS endmembers ($\mu_k$) for underlying biomolecules within an unlabelled dataset ($y_n$) by fitting a GMM with a prior estimate of the number of biomolecules ($k$). This method relies on three conditions: (1) target biomolecules must be abundant in the dataset, (2) each target biomolecule must provide a unique PARS signal, (3) a reasonable estimate of the number of biomolecules ($K$) must be available. Under these conditions, a biomolecules PARS signature or endmember can be extracted from the unlabelled data through GMM fitting, avoiding any requirement for a ground truth such as purified biomolecules, or chemically stained images (i.e., HC or IHC).

Assuming the endmember biomolecules are common within the dataset (i.e., most pixels belong to an endmember class), the $k$ different gaussian distributions will converge to the underlying distributions which represent the independent

Photon Absorption Remote Sensing (PARS)biomolecule signals, with some allowance for noise and overlap between distributions. The PARS signatures for each biomolecule are subsequently extracted as the expectation or mean ($\mu_k$) of the $k$ fitted distributions in the GMM.

An iterative expectation maximization (EM) method is used to fit the GMM,[73] estimating the endmembers distribution mean ($\mu_k$), covariances ($\Sigma_K$) and prevalence ($\Pi_K$) from the input data (Figure 3). The EM method guarantees convergence, but not which local minima the solution converges to. Hence, the EM is intentionally initialized with prior estimates of the mean, covariances, and prevalence, which are calculated from semantic labelling of the data to guide GMM convergence. An example of the GMM fitting process is shown in Figure 3a, for PARS signals corresponding to a section of human skin tissues. For illustration purposes, the 6-dimensional dataset is projected into 2-dimensional space (Figure 3a-1), comparing the non-radiative decay rate at $266\ nm$ and $532\ nm$. The simplified projection in the contrast shows three distinct signal clusters which are used to illustrate the principles of the GMM fitting. In practice, for labelling purposes, the raw data was contrast adjusted and projected to maximize visual differentiation of relevant biomolecules within the specimen. A consulting clinician then manually annotated relevant biomolecular features within the enhanced tissue images (Figure 3a-2), loosely corresponding to the three labels shown in the contrast domain. The estimate or initial values of the mean ($\mu_k$), covariances ($\Sigma_K$) and prevalence ($\Pi_K$) were then calculated from the cluster membership assigned in the manual labelling step and used to estimate the posterior distributions.

Once seeded, the GMM was fitted using the EM algorithm (Figure 3 b). The expectation step calculates the membership probabilities using ($\mu_k, \Sigma_k, \Pi_k$) from the posterior distribution. Using the resulting probabilities, the maximization step recalculates a new ($\mu_k, \Sigma_k, \Pi_k$). The steps are repeated iteratively, until the mixture model parameters stop changing significantly. The converged signal mean values $\mu_k$ corresponding to the gaussian centroid means were extracted as the expected PARS signal endmember or signature for each of the $k$ given biomolecules (Figure 3a-3). The estimated biomolecule endmembers ($\mu_k$) were then used to invert the linear mixture model, with the aim of determining the abundance ($a_{nk}$) of biomolecules at each pixel ($n$) based on the PARS measurements at each pixel ($y_n$). For each pixel, the $k$ biomolecule abundances were estimated from the PARS endmembers ($\mu_k$) through non-negative least squares unmixing (Figure 3a-4). In the presented example, the background feature (shown in dashed red in Figure 3a-3) is colored in black to ease viewing.

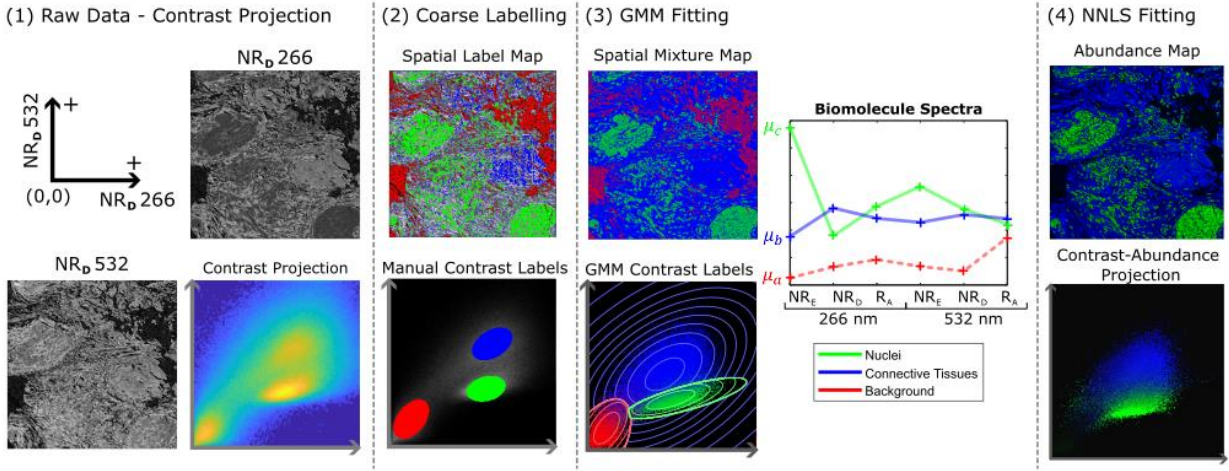

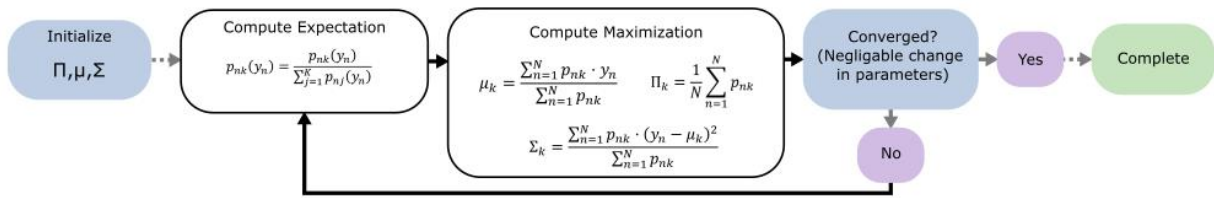

**Figure 3:** Process to extract biomolecule abundance estimations using gaussian mixture models (GMM),[73] and non-negative least squares (NNLS). (a) Workflow to identify and extract PARS endmembers or signatures, and abundances from PARS data. The data for this example is contrast projected into a 2-dimensional space for viewing. Biomolecules are semantically labelled by clinicians to seed the posterior mean ($\mu_k$), covariances ($\Sigma_K$) and prevalence ($\Pi_K$) values for GMM fitting. The coarse labelling in this example highlights regions of tissue which are differentiable in both the contrast, and spatial domain projections. Following GMM fitting (b) the resulting centroid mean values ($\mu_k$)

are extracted as the $k$ biomolecule endmembers or PARS signatures. Abundance mapping is calculated by inverting the linear mixture model using the PARS endmembers from the GMM. Unmixing is performed using the NNLS method, where color intensity in the resulting image corresponds to biomolecule abundance. Note: the background endmember is projected as black for clarity of viewing. (b) Process of expectation maximization used to fit the GMM to the n-dimensional PARS data. The GMM is initialized using semantic labeling to define the number of biomolecules $k$, and the parameters $\mu_k$, $\Sigma_K$ and $\Pi_K$. The expectation or membership probability $p_{nk}$ is calculated for each measurement $y_n$. Based on the expectation, the parameters $\mu_k$, $\Sigma_K$ and $\Pi_K$ are updated. These steps are repeated iteratively until there is a negligible change in the parameters $\mu_k$, $\Sigma_K$ and $\Pi_K$ during the maximization step.

### E. Sample Preparation

The formalin fixed paraffin embedded human skin tissues were provided by clinical collaborators at the Alberta Precision Laboratories (Calgary, Alberta, Canada) from anonymous patient donors. Samples were fully anonymized, and no patient information or identifiers were provided to the researchers. Patient consent was waived by the ethics committee under the condition that samples were archival tissues not required for patient diagnosis. The formalin fixed paraffin embedded murine brain tissues were provided by Dr. Deepak Dinakaran from the Sunnybrook Research Institute at the University of Toronto (Toronto, Ontario, Canada) under AUP #25997.

For both sample types, the tissue preparation protocol uses standard clinical practice guidelines. In general, sample tissues are placed in 10% neutral buffered formalin within 20 minutes of resection and fixed for 24 to 48 hours. Post fixation, tissues are dehydrated using a series of alcohols, then cleared with a xylene rinse. Cleared tissues are embedded in paraffin wax to produce formalin fixed paraffin embedded (FFPE) tissue blocks. Thin tissue sections (~4-5 µm) are cut from the FFPE blocks and placed on a microscope slide. Unstained sections are transported to the PhotoMedicine Labs at the University of Waterloo for imaging. Slides are briefly heated to 60ºC to smooth the paraffin sections prior to scanning. Once PARS imaging is completed, tissue sections are stained with the dyes (e.g., H&E) and sealed with a cover slip. Stained sections are imaged using a 40x digital pathology scanner (MorphoLens 1, Morphle Labs) to produce matched PARS and stained images.

These study operations were performed in accordance with ethics protocols approved by the Research Ethics Board of Alberta (Protocol ID: HREBA.CC-18-0277), Sunnybrook Research Institute (AUP: 25997) and the University of Waterloo Health Research Ethics Committee (Photoacoustic Remote Sensing (PARS) Microscopy of Surgical Resection, Needle Biopsy, and Pathology Specimens; Protocol ID: 40275). All human tissue experiments were conducted in accordance with the government of Canada guidelines and regulations, such as "Ethical Conduct for Research Involving Humans (TCPS 2)".

### F. Model Training

Training for the Pix2Pix model was carried out in a similar fashion as reported by Boktor et al.[52] Compared to this work, two main modifications were made to the model. A Resnet generator was implemented, and the model was expanded to accept 6 input channels corresponding to the radiative amplitude ($R_A$), the non-radiative energy ($NR_E$), and decay rate ($NR_D$) for each of the 532 nm and 266 nm excitation.

Model training data was produced by registering PARS images of murine brain tissues with the corresponding matched H&E from the same tissue section as outlined by Boktor et al.[52] Registered images were diced into 512 x 512-pixel patches, resulting in approximately 3000 training pairs (where the data presented in this paper was specifically excluded from this set). The resulting model was trained over 200 epochs, with a batch size of 1, where the dataset was split 85:15 for training and validation. All other initial model parameters and weights remained the same as described by Boktor et al.[52] To produce virtually stained images, the trained model was repeatedly applied to overlapping 512 x 512-pixel patches of unseen PARS data. Patches were diced with 75% overlap, and the resulting colorization was averaged together at each pixel to avoid artifacts present at the edges of adjacent colorization patches.

## IV. Results & Discussion

### A. Application to Thin Sections of Preserved Human Tissues

The presented PARS microscope uses a transmission mode design optimized for imaging thin histopathology samples (thin translucent sections of formalin fixed paraffin embedded tissues). Selected specimens underwent standard clinical sample preparation, where excised specimens were fixed, embedded into paraffin, sectioned, and mounted on slides. Unstained thin tissue sections were imaged with the PARS microscope, then returned for chemical staining according to clinical standards. Through this process PARS data is collected without impacting the standard clinical workflow, while allowing contrast multiplexing where PARS and chemical staining are performed on the exact same specimens for one-to-one imaging comparisons. As an example, a PARS scan capturing an excised human skin tissue sample exhibiting Basal Cell Carcinoma is presented in Figure 4, while the corresponding chemical H&E from this section of tissue is presented in SI: Figure 1.

# Photon Absorption Remote Sensing (PARS)

The measurements captured in PARS characterize most facets of de-excitation from an electronic absorption interaction, aiming to maximize the biomolecular contrast. Resulting PARS measurements of the non-radiative energy ($NR_E$), decay rate ($NR_D$), and radiative amplitude ($R_A$) are presented in Figure 4. All contrasts, from both 266 nm and 532 nm excitation wavelengths, are overlaid in Figure 4(a) emphasizing the full depth of data collected during a single PARS acquisition. The array of contrasts corresponds to an increase in dimensionality, and an increase in the capacity for biomolecule selectivity from each excitation, with unique biological features found from each pump wavelength, as compared to alternative independent methods (e.g., auto-fluorescence, photothermal, photoacoustic). The only absorption characteristics excluded in the current embodiment are the radiative emission spectra and lifetime, which may be explored in future works.

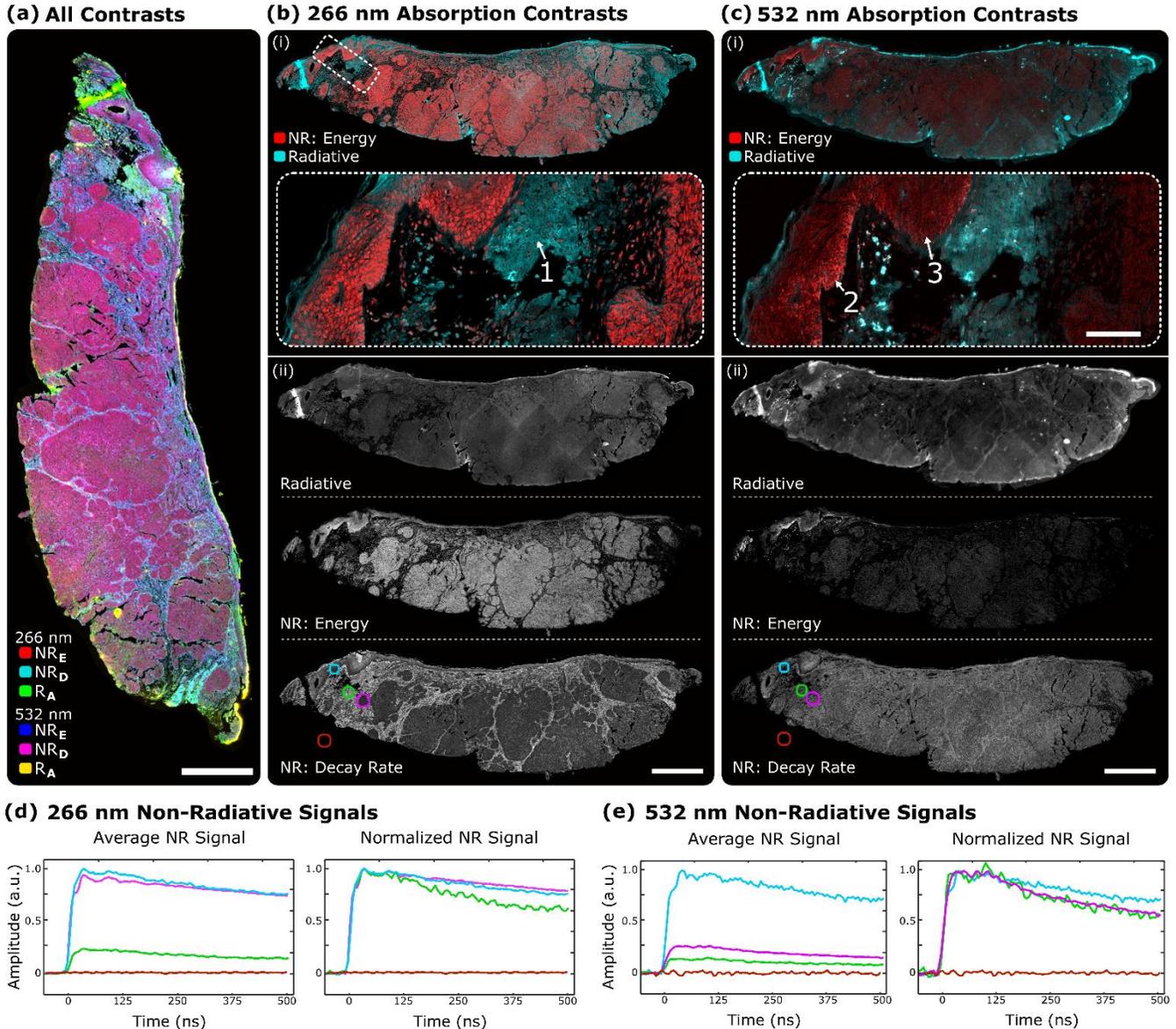

**Figure 4:** PARS image of a thin section of FFPE human skin tissue exhibiting basal cell carcinoma. (a) PARS image, showing all absorption contrasts from both 266 nm and 532 nm excitation merged into a single image. Scale Bar: 500 µm. Enhanced subregion presented to show fine detail contrast of the TA images. Scale Bar: 100 µm. (b-i) TA PARS image showing the merged non-radiative energy, and radiative amplitude at the 266 nm wavelength. (b-ii) Individual absorption contrasts captured in the same skin tissue section as (a), for the 266 nm excitation wavelength. Images show the radiative amplitude, non-radiative energy, and non-radiative decay rate, where rime domain signals from each highlighted ROI are presented in (d). Zoomed in sections of the non-radiative decay rate images are presented in SI: Figure 2. Scale Bar: 500 µm. (c-i) TA PARS image showing the merged non-radiative energy, and radiative amplitude at the 532 nm wavelength. (c-ii) Individual absorption contrasts as presented in (b) for the 532 nm excitation wavelength. Scale Bar: 500 µm. (d) Average unfiltered time domain signal, from the 266 nm excitation corresponding to the regions marked in (b). Sharper decay values correspond to brighter pixels in (b). (e) Unfiltered time domain signals from the corresponding regions in (c).

In the skin tissues, the deep UV excitation (266 nm) is broadly absorbed by several relevant biomolecules including collagen, elastin, melanin, nuclei and DNA. The visible (532 nm) excitation also targets multiple common bioproteins (e.g., collagen, elastin, hemoglobin), providing a secondary measurement of similar features highlighted by the UV. Resulting visualizations from each excitation are shown in Figure 4 (b) and (c), where panel (i) shows the PARS TA image, while (ii) shows the independent channels. The TA image characterizes the complete relaxation of each pixel, showing the total energy shed through both non-radiative (red), and the radiative (blue) de-excitation phenomena. These TA images (Figure 4 (b-i) and (c-i)), aid in emphasizing the significant differences in relaxation characteristics between absorbers. For instance, in the 266 nm excitation the non-radiative contrast (Red: Figure 4 (b)) highlights predominantly nuclei, as DNA exhibits a low QY. The radiative contrast (Blue: Figure 4 (b)) emphasises biomolecules with higher QY (e.g., keratin, elastin).

By measuring absorption as the combination of both dominant relaxation effects, PARS decouples the absorption contrast from mechanism specific efficiency factors such as the QY or PCE. In principle any biomolecule which absorbs light will offer some degree of PARS contrast (either radiative or non-radiative). This is especially relevant in cases where the contrast in a region of tissue is dominated by one relaxation effect. For instance, consider region (1) in Figure 4 (b-i). The connective tissues around label (1) exhibit near complete radiative (blue) relaxation (Figure 4 b(i)). Capturing only non-radiative relaxation (Figure 4 b(ii)) would indicate low UV-absorption in this area and could even miss this region of connective tissues completely. In contrast, by measuring both relaxation fractions to provide the "Total-Absorption" (Figure 4 b(i)) PARS reveals the total level of UV absorption, independent of the contrast specific efficiency factors (e.g., QY, PCE).

Capturing the interplay of relaxation phenomena also provides a unique avenue of contrast, as the relative distribution of de-excitation (radiative vs. non-radiative) can reveal insights into biomolecules QY or PCE. Previously a PARS specific metric was proposed as the "quantum efficiency ratio" (QER) to characterize the relative de-excitation distribution (SI: Figure XX.).[53] The QER is calculated as: $(R_A - NR_E)/(NR_E + R_A)$, and scales from -1 for completely non-radiative to 1 for perfectly radiative PARS contrast. This relative measure is correlated with, but not equivalent to the QY or PCE, as it is subject to the limitations of the $NR_E$ and $R_A$ measurements, where factors such as temperature, mechanical properties, or optical architecture may change the observed contrasts. An example of the QER contrast in the tissue section presented in Figure 4, is shown in SI: Figure 3. In practice, the QER can be viewed through the color (red vs. blue) mixing of the TA image pixels, which can aid in differentiating biomolecules that appear structurally similar otherwise. An example is highlighted in the 532 nm images (Figure 4(c)), where both points (2) and (3) indicate regions of the basal layer (at the edge of the epidermis), which appear similar in the radiative image (Figure 4 c(ii)). However, in the TA image (with the QER coloring, Figure 4 c(i)), region (2) appears red orange indicating more non-radiative relaxation in this region. This corresponds to the presence of melanin in the basal layer of region (2), which is less prevalent in region (3).

As a further source of contrast PARS also extracts the time-evolution of the unfiltered non-radiative signals profiles. As outlined in the **_PARS Mechanism_** section, the non-radiative perturbations temporal evolution is largely dictated by thermal diffusion in the presented embodiment. To this end, a thermal decay constant (correlated to the local thermal diffusivity and volume of excited tissue) is extracted from the non-radiative modulation through a log-linear fitting. Normalized thermal decay rates across the tissue are presented in Figure 4(b) & (c) for each excitation, revealing the relative difference in decay properties across the specimen, which presumably relates to local tissue density, cellularity, and biomolecular composition. Darker pixels indicate a lower magnitude of decay rate (closer to zero). For reference, averaged non-radiative perturbations are presented in Figure 4 (d) & (e) for the regions highlighted in the absorption images. Enhanced images of the non-radiative time decay images are presented in SI: Figure 2. Small differences are observed in the normalized non-radiative perturbations, which are attributed to variance in the material properties and volume of tissue excited at each pixel. In this example, the local material properties are largely dominated by the paraffin wax substrate which the tissues are embedded in corresponding to very small changes in the decay rate across the specimen. However, there are appreciable differences between the highlighted regions indicating that the non-radiative decay rate offers an independent avenue to capture additional absorber parameters further increasing the contrast and specificity.

## B.     *Biomolecule Unmixing*

### i.    *Tissue Samples*

The presented method is initially applied to unmixing biomolecules within murine brain tissues with xenograft primary brain tumour (glioblastoma) implanted (Figure 5). A series of four subsections from various regions of the brain including the dense tumor (iv), the peripheral margin along the pia matter (ii), along the white matter tracts (iii), and within the center of the sample (i). As an initialization step, clinicians semantically labelled tissue features within the tissues, where the contrast was adjusted to emphasize clinically relevant structures. Some example annotations are shown in Figure 5. As an example, nuclei are characterised by a pink or fuchsia color, while grey matter exhibits a darker purple, finally white matter and red blood cells exhibit varying shades of green to cyan. These selections defined the initial GMM conditions (e.g., gaussian means ($\mu_k$)) serving to guide convergence towards the target biomolecules. The seeding could be skipped, however, the GMM is not guaranteed to converge on the desired clinical features if randomly initialized. Once the GMM is fitted, the gaussian mean vectors ($\mu_k$) are extracted as the PARS endmember or signature for each biomolecule.

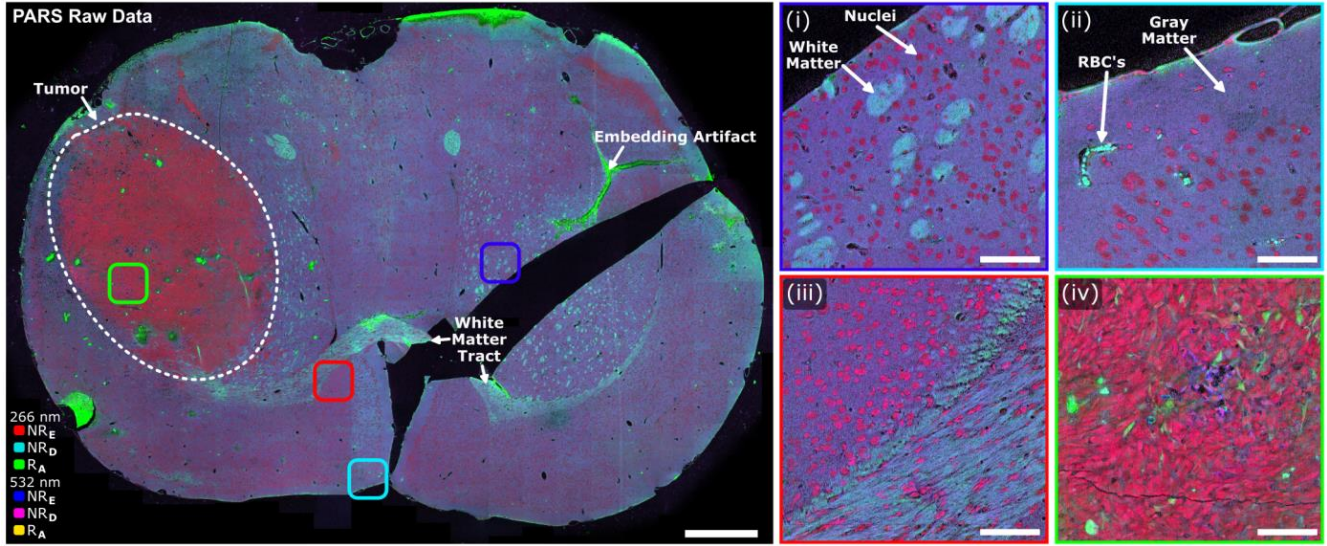

**Figure 5:** PARS image of thin sections of FFPE murine brain tissues. The PARS image, overalays all absorption contrasts from both $266\ nm$ and $532\ nm$ excitation merged into a single image. Scale Bar: $1\ mm$. Subregions shown in (i)-(iv) show fine detail contrast of the TA images. Scale Bar: $100\ \mu m$. (i) transverse cross section of white matter tracts, (ii) vessels and gray matter near the pia matter, (iii) dense white matter tracts, (iv) dense hyper-nuclear tumor tissue.

### ii.    Statistical Unmixing Results

Based on the initial seeding in this sample, a total of five unique molecules were identified through the GMM fitting including nuclei, white matter, gray matter, red blood cells, and paraffin wax. The corresponding PARS signatures produced by the GMM fitting are shown in Figure 6. The PARS endmembers are normalized by the global average for each contrast channel, meaning signatures should be interpreted as the contrast for a specific biomolecule relative to the global average. For instance, the 266 nm $NR_E$ for nuclei (Figure 6(b)) is more than double the global average. The PARS signatures also illustrate the relative distribution of relaxation characteristics expected from a given biomolecule. As an example, based on the PARS signature observed in Figure 6(d), white matter tends to exhibit higher $R_a$ at 266 nm, and $NR_E$ at 532 nm, while the other relaxation effects are much lower.

Biomolecule abundances are then estimated by non-negative least squares (NNLS) unmixing of the endmembers. Abundance mappings are produced (Figure 6(i)-(iv)) for the sub-regions of the brain tissues shown in Figure 5. The top row (Figure 6(a)) presents the overlaid abundance mappings from all endmembers except the paraffin wax, which is not shown to emphasize the diagnostically relevant biomolecules. Within these complex samples, nuclei (green) are clearly differentiated and identified from the surrounding gray matter (blue). Even small volumes of gray matter are discerned in sections of dense tumor tissue largely composed of nuclei (Figure 6(iv)). In other sections near the corpus callosum (Figure 6(i) and (iii)), white matter tracts (pink) are clearly delineated in both transverse (Figure 6(i)) and longitudinal (Figure 6(iii)) orientation within the tissues. Red blood cells are also observed in the microvasculature of the gray matter near the periphery of the brain (Figure 6(iii)), and in the dense tumor tissues (Figure 6(iv)). Finally, paraffin wax is identified along the edge of the tissues in (Figure 6(i) and 6(ii)), and within the holes present in all tissue examples. This directly provides contrast to critical biomolecules and tissue features required for histopathological assessment of tissues in clinical settings.

While the presented GMM and NNLS abundance mappings exhibit exceptional separability between biomolecules within the complex brain tissue samples, there are limitations. The performance of the NNLS abundance mappings is imperfect due to noise within the individual PARS measurements at each pixel, potential similarity between endmembers, and imaging system parameters (e.g., difference in resolution between contrasts). This is most prevalent in the RBC (Figure 5 (e)), and white matter endmembers (Figure 5 (d)). An additional limitation is that extracted endmembers may not be transferrable between samples. As an example, the nuclear endmember may differ slightly between the murine brain tissues (Figure 5) and the previous skin tissue sample (Figure 4), due to differences in the tissue type, paraffin substrate, or imaging parameters. This challenge can be mostly circumvented as the endmember extraction and unmixing is performed directly on the presented data. With some external seeding input, the same GMM and NNLS approach could be applied to extract a nuclear (or other biomolecule) endmember within the skin tissues, or any other tissues samples. This includes specimens where capturing a ground truth is inconvenient, difficult, or impossible (e.g., freshly resected tissues), offering a clear advantage over paired methods.

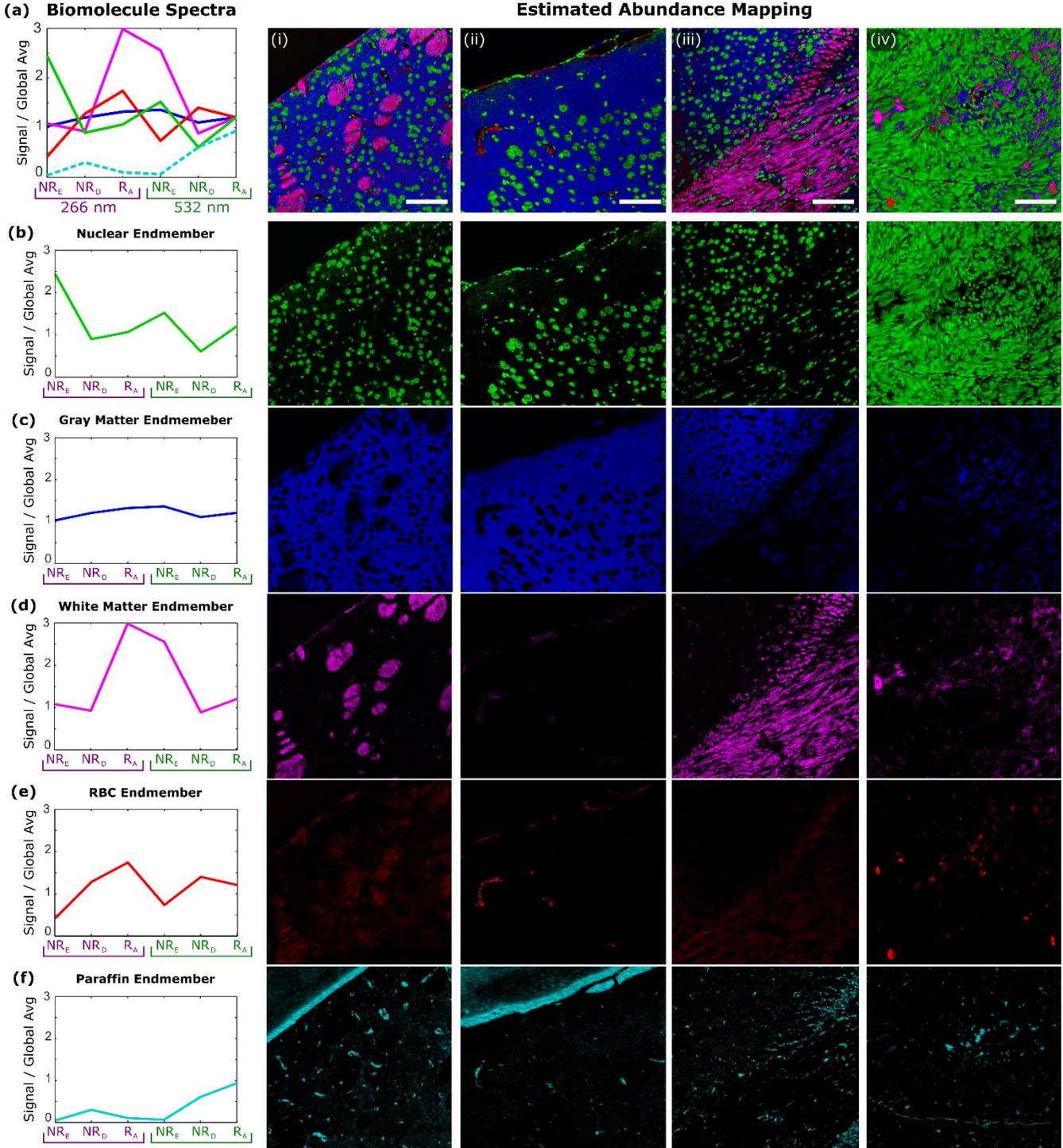

**Figure 6:** PARS biomolecule endmember profiles and abundance mappings. (a) Biomolecule endmembers produced from GMM fitting of PARS data, and several examples of abundance mapping in different regions of FFPE murine brain tissues. Target biomolecules include Nuclei, Gray Matter, White Matter, Red Blood Cells, and Paraffin Wax. Note that paraffin wax abundance mapping is not presented in (a) to better emphasize clinically relevant biomolecules. Corresponding sections in (b) to (f) show the isolated endmembers and abundance mapping for the (b) Nuclei, (c) Gray Matter, (d) White Matter, (e) Red Blood Cell, (f) Paraffin wax. (i)-(iv) Scale Bar: 100 $\mu m$.

*iii.    Nuclear Differentiation*

PARS facilitates this robust biomolecule specificity by capturing the relative distribution of relaxation effects following the absorption interaction. In most cases, each independent relaxation characteristic (e.g., non-radiative energy, radiative energy), may not provide sufficient selectivity to reliably differentiate biomolecules. This is especially challenging in complex samples, where the wavelengths (UV to NIR) used to generate excited state transition inevitably excite numerous biomolecules. For example, photoacoustic works often use UV (266 nm) to target nuclei,[56] however, many biomolecules provide non-radiative contrast following 266 nm excitation. In the brain tissue specimen (Figure 7), the $NR_E$ at 266 nm (Figure 7(b) or red in Figure 7(a)), shows appreciable contrast from nuclei and surrounding gray matter elements, which may impede reliable discrimination of nuclei. In contrast, by extracting the nuclear abundance with the presented method (Figure 7(d)), a nuclear mapping can be generated revealing isolated nuclei dispersed throughout the tissues. This approach supresses confounding signals from the surrounding gray matter. As a ground truth the corresponding DAPI (Figure 7(d)), and H&E (Figure 7(e)) staining within the same tissue specimen are also presented. The H&E highlights the nuclei in purple, while also confirming the presence of gray matter features surrounding the nuclei (pink). The DAPI stain specifically binds to DNA in nuclei, confirming the distribution of isolated nuclear structure dispersed throughout the gray matter. Hence the proposed method, exhibits how PARS native label-free contrast can begin to approach the specificity of the chemical (e.g., DAPI, H&E) staining by capturing and characterizing a complete representation of the de-excitation characteristics of biomolecules.

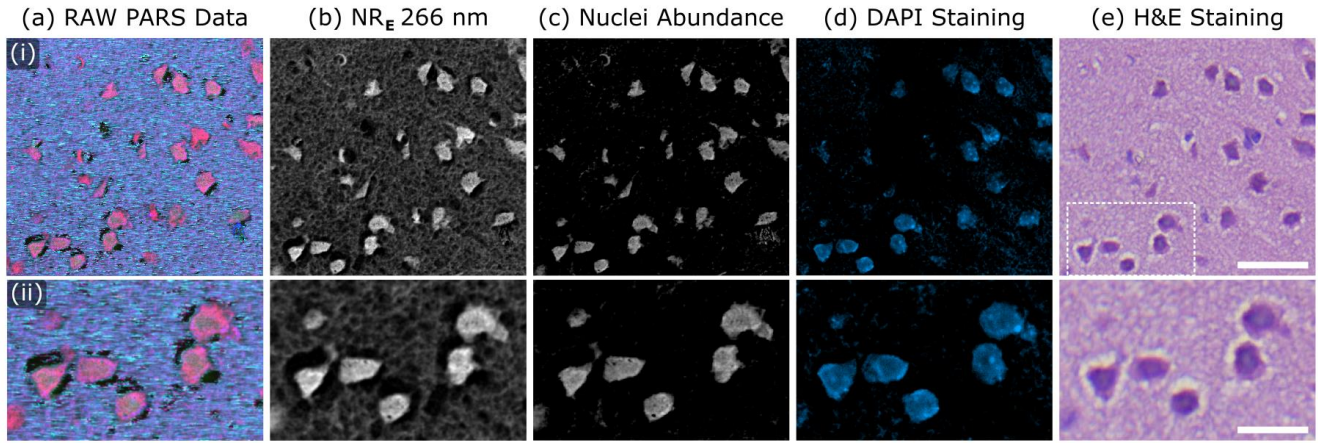

**Figure 7:** Comparison of nuclear contrast in the (a) the full Raw PARS contrasts, (b) the $NR_E$ at 266 nm commonly used to target nuclear features.[56] (c) the Nuclear Abundance mapping derived from the proposed unmixing method, and chemically stained (d) DAPI, and (e) H&E ground truth images in a section of murine brain tissues. The raw PARS data (a) contains the full suite of 6 contrasts ($NR_E, NR_D, R$, from both excitations), while the $NR_E$ isolated the channel typically used in label-free absorption methods to represent nuclear abundance. The Nuclear Abundance (c) uses the endmembers shown in Figure 5 to perform NNLS unmixing. Scale Bar (i) 50 $\mu m$, (ii) 25 $\mu m$

## C.    *Emulating Histochemical Contrasts: Virtual Staining*

As a further exploration, the presented biomolecule unmixing workflow is also applied to virtual staining. Virtual staining is an emerging method aiming to replace chemical staining with label-free microscopy and deep-learning. In virtual staining, label-free data is matched to stained images, and an image-to-image translation network (e.g., Pix2Pix,[74] cycleGAN[75]) is trained. The network learns correlations between the structures and contrasts of the label-free and chemically stained image domains. Trained networks can then estimate the chemical staining visualization for a label-free input.[19,76] While deep-learning based virtual staining has gained popularity in recent years, there is apprehension around the reliability of virtual staining.

Virtual staining performance is ultimately driven by the contrast and structural content of the label-free input, where deficiencies in biomolecule specificity can cause unexpected outcomes. For instance, poor label-free specificity can increase models' dependence on spatial context of tissue, resulting in sensitivity to subtle morphological differences. In the worst case, this can lead to colorization instability and critical hallucination errors (i.e., convincing structures are generated in the virtually stained image that do not exist in the ground truth).[19,76] This challenge may become insurmountable when attempting to generalize algorithms to perform predictably and reliably on all ranges of unseen data, as all possible examples must be sufficiently represented in the training set to avoid unforeseen consequences. These concerns have plagued absorption modalities like autofluorescence and photoacoustic microscopy and led to legitimate apprehension on the reliability of "black box" approaches in clinical settings.

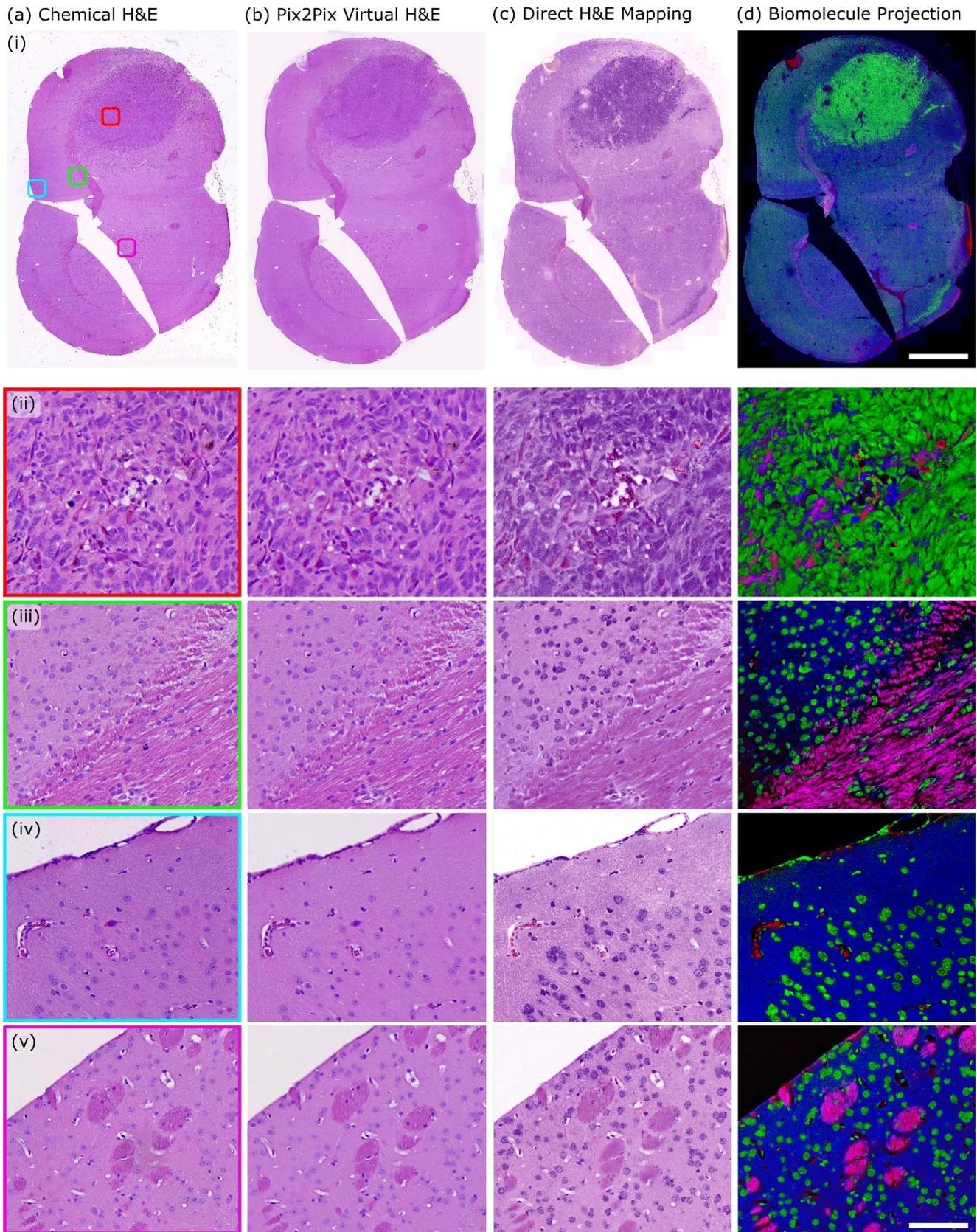

**Figure 8:** Comparison of PARS visualizations, and chemical H&E staining across sections of murine brain tissues. (a) Raw PARS data, presented in the same color mapping as shown in Figure 4. (b) GMM and NNLS biomolecule abundance mapping, highlighting the same biomolecules as shown in Figure 5. (c) Direct H&E mapping, by re-coloring the abundance estimates to match the colors provided by H&E staining. (d) Pix2Pix based virtual staining of the same tissue section, using the Raw PARS data as input. (e) Chemical or ground truth H&E-stained image, scanned using a brightfield transmission microscope. (i) Scale Bar: $2\ mm$. (ii) Scale Bar: $100\ \mu m$.

To explore the potential of PARS for virtual staining, the murine brain tissue specimen (Figure 5) is virtually H&E stained and presented in Figure 8(b). The virtual staining is performed using a Pix2Pix deep-learning model as described by Boktor et al.[52] As a ground truth comparison, the brain tissue sample was chemically H&E stained, then imaged using a brightfield microscope (Figure 8(a)). In each of the H&E representations, (Pix2Pix virtual H&E, and Chemical H&E) the same diagnostic features are identified. In the whole slide image, the dense region of tumor is visible in the top of the tissue sample, while white matter tracts are observed across the left middle. In the selected subregions, the morphology and cellular level distribution of individual nuclei is readily discerned (Figure 8(ii-vi). Concurrently, white matter tracts are observed in dark pink spreading throughout the gray matter (Figure 8(iv) & (vi)). Finally, red blood cells within the brain microvasculature (Figure 8 (v)) and tumor tissue (Figure 8 (ii)) are similarly identified in each image.

As evidence of the PARS virtual staining performance, we circumvent the (Pix2Pix) deep-learning network and instead replicate the H&E staining contrast using the proposed GMM and NNLS method. The same PARS endmembers extracted in the ***Biomolecule Unmixing*** section (Figure 6) are used for this process. In practice, the whole slide PARS image of the murine brain tissues (Figure 8(d)) was processed to map the abundance of nuclei, white matter, gray matter, and red blood cells. The corresponding abundance estimates (Figure 8(d)) are then re-colored to match the H&E staining palate. The nuclei appear purple, the white and gray matter are colored in different shades of pink, and the red blood cells appear in shades of red. The resulting images are presented in Figure 8(c) as the Direct H&E Mapping. Notably, this method does not use spatial context, as each pixel is processed independently (i.e., NNLS estimates abundance from signal and assigns H&E color).

In the presented results, the virtual staining differs from the Direct H&E as the deep-learning algorithm can leverage the structural context, adaptively suppress or enhance data, and provide denoising functionality. In addition, compared to the Direct H&E method, the Pix2Pix workflow transforms the PARS data to match the appearance of the chemical H&E. The appearances differ as the ground truth H&E was imaged using a brightfield transmission microscope, while PARS observes transient absorption effects. Despite these differences, the presented results are a critical first step towards developing trustworthy virtual staining models, by showing that PARS can provide the underlying specificity to replicate chemical stains (specifically H&E). With the presented GMM and NNLS workflow, the diagnostic biomolecules stained by H&E dyes are reliably characterized and mapped from PARS signals alone, while avoiding reliance on structural context. This results in effectively equivalent contrast between the virtual H&E, the PARS Direct H&E, and the chemical H&E (Figure 8).

## V. Conclusions

In principle, this work provides the first comprehensive explanation of PARS, a promising new optical absorption microscopy technique. The complete PARS mechanism is explored illustrating the comprehensive array of contrasts, which are derived directly from endogenous chromophores. The combination of measurements captured in PARS are designed to characterize most facets of de-excitation from an electronic absorption interaction, with the aim of maximizing biomolecular contrast. The resulting PARS microscope directly combines the strengths of radiative (e.g., autofluorescence), and non-radiative (e.g., photothermal, and photoacoustic) techniques into a single modality. For example, capturing the absorption as the sum of relaxation fractions may circumvent contrast specific efficiency factors (e.g., QY) common to alternative methods. Concurrently, viewing the de-excitation phenomena as coupled interactions, enables characterization of biomolecules based on their distribution of de-excitation effects. In combination with the material property measurements captured through the non-radiative signal temporal evolution, PARS may facilitate an enhanced recovery of a wider range of biomolecules than independent radiative or non-radiative modalities from each excitation event.

This provides a unique opportunity to take on some of the predominate challenges impeding label-free absorption microscopy for histopathological imaging of tissues. Using well-known statistical processing methods (GMM and NNLS), PARS can extract unique signatures characterizing clinically relevant biomolecules, then unmix these biomolecules within complex media to produce spatial abundance mappings. The PARS unmixing and abundance estimates show high structural similarity, and accuracy when compared against chemically stained ground truth images, and deep learning based-image transforms. Moreover, these results show PARS capacity to directly image and unmix the critical biomolecules necessary for clinical diagnostics and histopathological assessment of tissues. This has particular relevance in disease processes such as brain tumours which routinely require molecular staining and pathologic testing over-and-above H&E-based tissue morphology assessment by a clinician. While this study has a limited dataset, these findings strongly support PARS previous successes with deep learning-based image transforms,[51,57] as established statistical methods can provide biomolecule specificity without a chemically stained ground truth. This provides significant backing for future developments of reliable virtual staining workflows. Ideally, this can help to mitigate the distrust and opposition to "black box" approaches in clinical settings, by directly approaching the selectivity limitations of current label-free histopathology methods.

In future, the presented method may be applied to provide label-free biomolecule specific visualizations in more complex specimens, and cases where ground truth images cannot be produced (e.g., freshly resected tissues). Alternate specimens may feature unique diagnostic biomolecules encouraging further expansion of the PARS contrasts. For example, future works will explore capturing the radiative emission spectra, and signal lifetime. In addition, further excitation wavelengths may be integrated, each targeting a unique set of biomolecules. Overall, PARS represents an appreciable step towards a new paradigm of absorption microscopy where each absorption event is comprehensively characterized by the interplay of all relaxation effects. Moreover, with research moving rapidly towards deep learning-based processing, and AI driven diagnostics, this rich combination of absorption contrasts represents an exciting new source of data.

**Acknowledgments**
The authors thank Dr. Marie Abi Daoud at the Alberta Precision Laboratories in Calgary, Canada for providing the human skin tissue samples. Additionally, the authors would like to acknowledge Hager Gaouda for their valuable assistance in staining the tissue samples used in this study.

**Contributions**
B.R.E conducted experiments, formulated and applied process, and wrote manuscript; J.A.T.S helped conduct experiments and contributed to manuscript writing; J.E.D conducted experiments, and contributed to manuscript writing; D.D helped in collecting results, procuring samples, annotating images, and guiding experiments; P.H.R served as principle investigator;

**Funding Sources**
This research was funded by: Natural Sciences and Engineering Research Council of Canada (DGECR-2019-00143, RGPIN2019-06134, DH-2023-00371); Canada Foundation for Innovation (JELF #38000); Mitacs Accelerate (IT13594); University of Waterloo Startup funds; Centre for Bioengineering and Biotechnology (CBB Seed fund); illumiSonics Inc (SRA #083181); New frontiers in research fund – exploration (NFRFE-2019-01012); The Canadian Institutes of Health Research (CIHR PJT 185984), (PJT-195962).

**Competing Interests**
Authors Benjamin R. Ecclestone, James A. Tummon Simmons, James E. D. Tweel, Deepak Dinakaran, and Parsin Haji Reza all have financial interests in IllumiSonics which has provided funding to the PhotoMedicine Labs.

## VI. Supplemental Information

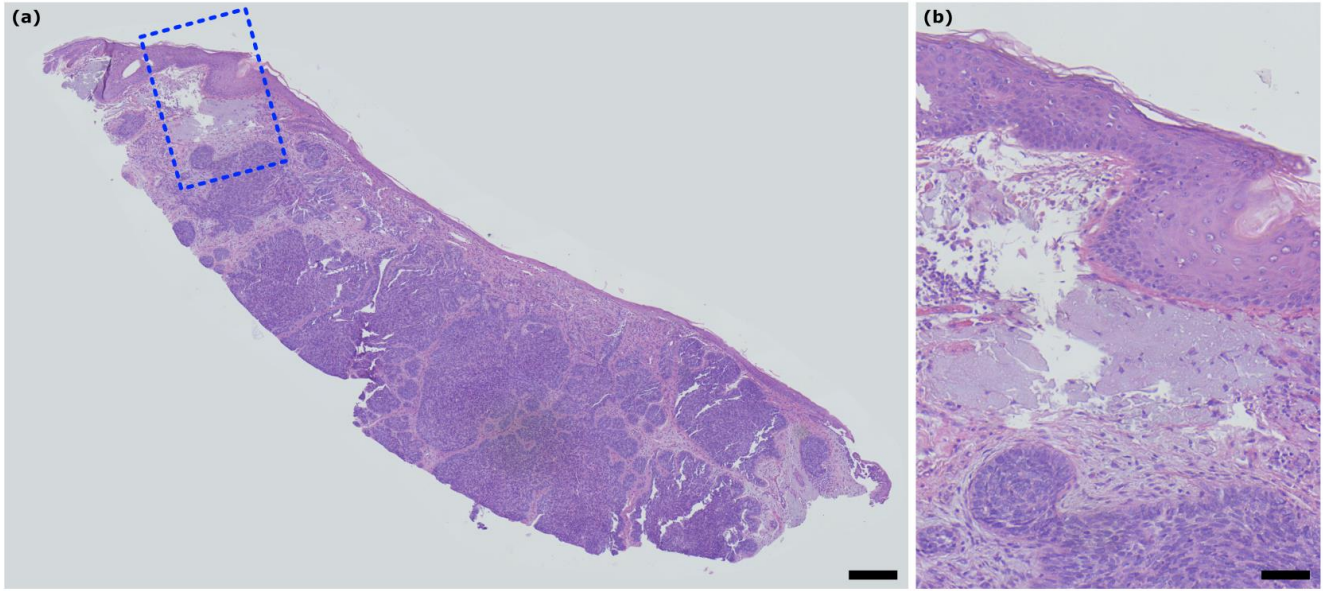

**Supplemental Figure 1:** H&E stained brightfield microscope image of a thin section of FFPE human skin tissue exhibiting basal cell carcinoma. (a) Whole slide image. Scale Bar: 250 $\mu m$. (b) Enclosed subsection of the same skin tissue section. Scale Bar: 30 $\mu m$.

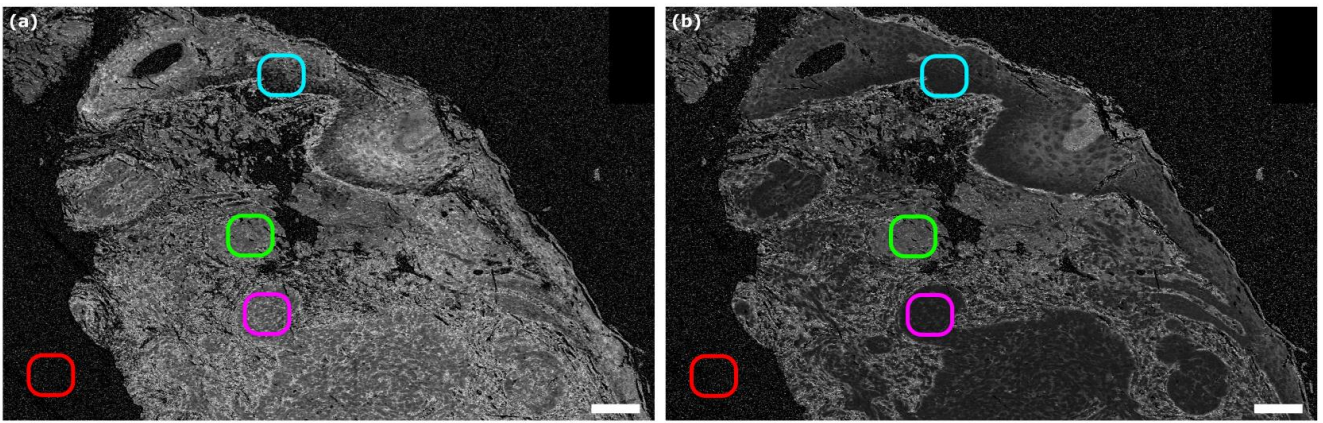

**Supplemental Figure 2:** Non-Radiative decay rate contrast in a thin section of FFPE human skin tissue exhibiting basal cell carcinoma. Enhanced view of the images shown in Figure 4, illustrating the non-radiative decay rate, where rime domain signals from each highlighted ROI are presented in Figure 4 (d). Scale Bar: 100 $\mu m$. (a) 532 $nm$ excitation wavelength. (b) 266 $nm$ excitation wavelength.

Photon Absorption Remote Sensing (PARS)

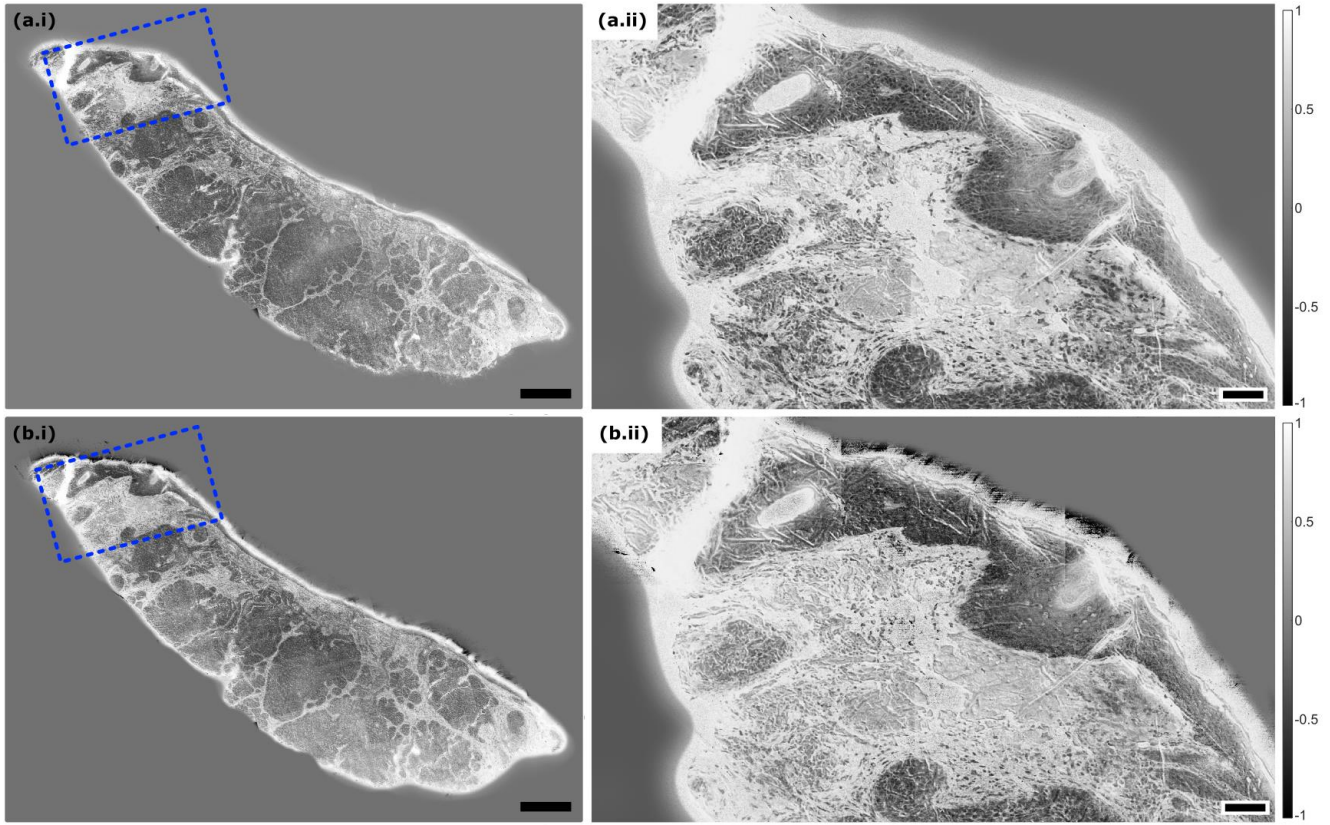

**Supplemental Figure 3:** PARS QER metric calculated as ($QER = (R_A - NR_E)/(NR_E + R_A)$) across the thin section of FFPE human skin tissue exhibiting basal cell carcinoma shown in Figure 4. (a) PARS QER image from the 266 nm wavelength. (b) PARS QER image from the 532 nm wavelength. (i) Whole slide image. Scale Bar: 100 $\mu m$. (ii) Encapsulated subregion showing an enhanced region. Scale Bar: 100 $\mu m$.

23